\documentclass[11pt]{article}
\usepackage{amsmath,amssymb,color,bbm}
\usepackage{color}
\usepackage{graphicx}
\usepackage{subcaption}
\usepackage{wrapfig}
\usepackage[percent]{overpic}
\usepackage{caption}
\usepackage{mathtools,xparse}
\usepackage{geometry}
\usepackage{float}
\usepackage{url}
\usepackage{verbatim}
\usepackage{lmodern}
\usepackage[square,sort,comma,numbers]{natbib}
\bibpunct{[}{]}{,}{n}{}{;}
\usepackage{hyperref}

\hypersetup{
	colorlinks=true,
	citecolor=blue,
	filecolor=blue,
	linkcolor=blue,
	urlcolor=blue,
	pdfstartview={FitR},
	filecolor=blue
}

\newcommand{\be}{\begin{equation}}
	\newcommand{\bea}{\begin{eqnarray}}
		\newcommand{\ee}{\end{equation}}
	\newcommand{\eea}{\end{eqnarray}}

\usepackage{lmodern}
\DeclareMathAlphabet{\mathbfsf}{\encodingdefault}{\sfdefault}{bx}{sl}

\usepackage[T1]{fontenc}
\usepackage[utf8]{inputenc}
\usepackage{authblk}

\title{Stability analysis of Euler's elastica ring using harmonic balance}
\author[]{Muhammad Sami Siddiqui \thanks{dr.samisiddiqui@iqra.edu.pk}}
\affil[ ]{FEST, Iqra University, Main campus Karachi, Pakistan}
\date{}

\begin{document}
		\maketitle
		\begin{abstract}
			\noindent	
			The stability analysis of elastic rings subjected to various loading conditions is examined, focusing on stable and unstable configurations. The harmonic balance method is employed to investigate the stability range under different loading conditions. This method provides improved accuracy in results and yields a smooth, stable range of values for all modes of symmetry. Additionally, the bifurcation analysis technique of linearization is utilized to identify unstable regions, and it demonstrates good agreement with the results obtained using the harmonic balance method.
		\end{abstract}
		Keywords: Euler's elastica ring; Harmonic balance method; Post buckling; Stability analysis; Linearization.
		
	\section{Introduction}
	
	The bending beam problem, first posed by Jacob Bernoulli in 1691 \cite{Bernoulli}, was partially solved by Leonard Euler \cite{Euler}, who formulated it in terms of stored energy and described bifurcation in the context of buckling. Lagrange \cite{Lagrange} made significant contributions to the elastica problem by considering slight deflections and deriving a second-order differential equation. He also obtained solutions for different buckling modes and critical load conditions. In 1811, A.J.C.B. Duleau \cite{Goss} conducted extensive experimental work on the stability of elastic beams. Max Born \cite{Born}, in his doctoral thesis in 1906, investigated the stability of planar and spatial elastica under various boundary conditions.
	
	The buckling of a circular elastic ring subjected to $n$-fold symmetric loading conditions and an external force $p$ has widespread applications in real-life problems, including biomechanics, structural stability, and soft robotics \cite{Hazel, CANHAM, Deborah, David, Tony, Elstc}. The study of buckling and post-buckling of elastic rings was initiated by Levy \cite{Levy} and Carrier \cite{Carrier}, and they have done detailed work on the problem. Tadjbakhsh and Odeh \cite{Tadjbakhs} rederived the formulation of the Euler elastic ring using the force-moment balance technique and obtained solutions for the buckled equilibrium shapes. They also established that the dimensionless pressure $p$ satisfies the condition $p_n > n^2 - 1$. Flaherty et al. \cite{Flaherty} numerically analyzed this problem and determined the contact and over-contact shapes for buckling modes $n = 2, 3, 4$.
	
	"Applying the Liapunov–Schmidt reduction in the
	bifurcation theory to the governing equation of the ring,
	Troger and Steindl ref and Chaskalovic and Naili ref also
	established the corresponding analytical approximate
	solution. These analytical approximate solutions are
	expressed in terms of power series of a small parameter
	but their range of validity is small. An interesting and
	robust numerical approach to this problem was proposed
	by Harrison using a discrete element and the shooting
	method"
	
	 Djondjorov et al. \cite{Djondjorov} provided an explicit formulation for the elastic ring subjected to a uniform symmetric external pressure of mode $n$ using elliptic functions and computed the contact points for different $n$. Wu et al. \cite{Baisheng} employed the harmonic balance method for mode $n = 2$ and compared it with other methods such as perturbation and numerical approaches. Majid and Siddiqui \cite{AMajid2} solved the Euler elastica problem with different approximation methods, considering a general formulation for $n$ rotational symmetry, and found that the third-order harmonic balance method performed better than the elliptic solution.
	
	When a physical system undergoes deformation due to the application of an external force, it bifurcates after reaching a critical value. These ranges define the stability of the system under a particular solution technique. This paper focuses on computing the stability range for the Euler elastica ring using the harmonic balance formulation under various loading conditions. This method and results are compared with the well-established methods and technique and made significant comparison. Additionally, the range illustrates the progression of ring deformation from stable to unstable shapes. The stability ranges are also computed numerically and compared with the harmonic balance results. Finally, the instability of the Euler elastica equation is analyzed using bifurcation theory and compared with previous findings.
	
	This paper is organized as follows. Section 2 discusses the two-dimensional theory of the elastic ring and the nonlinear differential equation governing its behavior. Section 3 introduces the harmonic balance method and solves the differential equation. In Section 4, stability ranges and elastic ring shapes are computed using the harmonic balance method for different loading values and conditions. These results are in agreement with the bifurcation theory of linearization. Numerical computations of stability ranges are also performed and compared with previous findings. Finally, Section 5 concludes the paper.

		\section{Model and Governing Equation}
			\begin{wrapfigure}{r}{6cm}
			\begin{center}
				\includegraphics[width=0.2\textwidth]{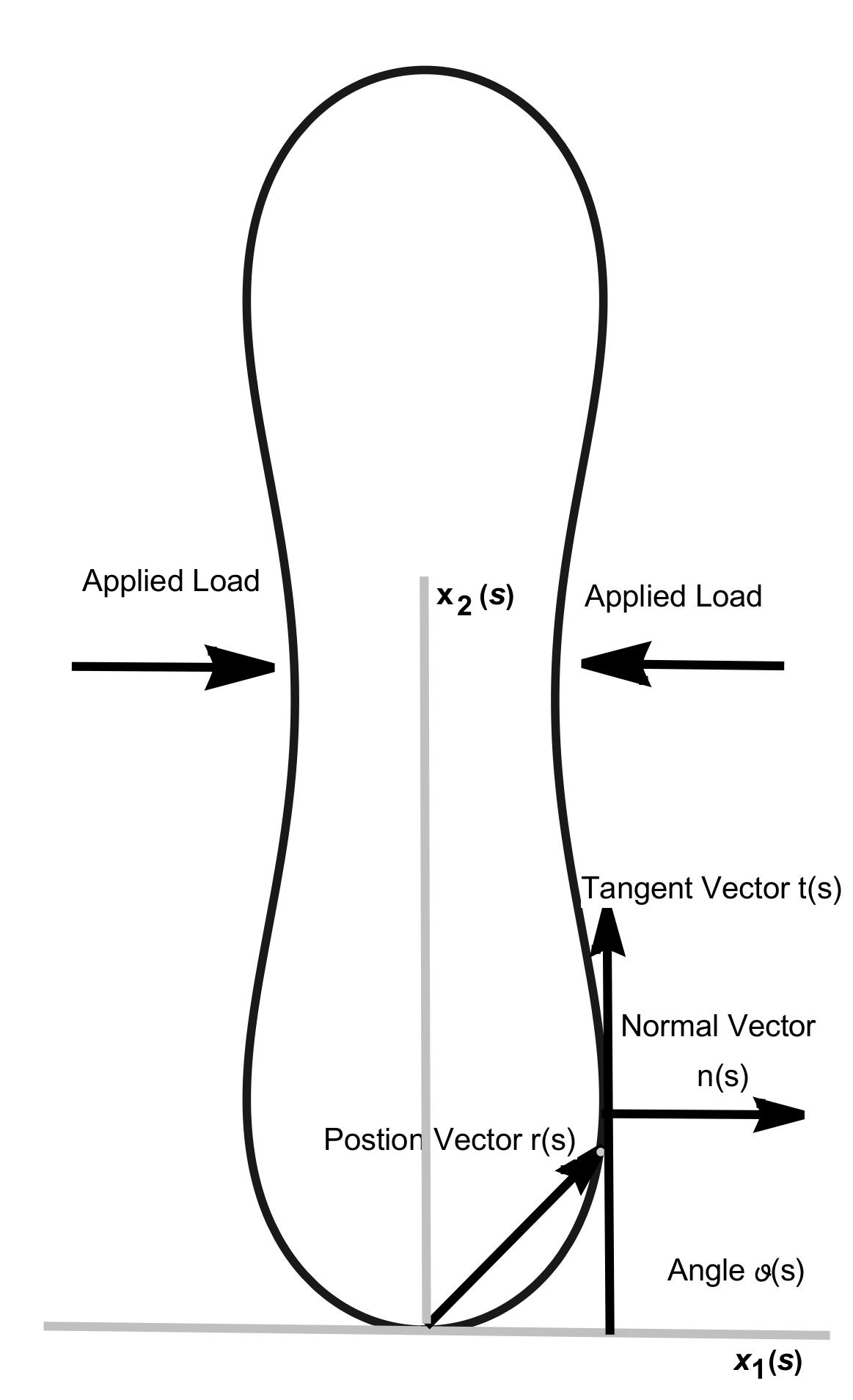}
			\end{center}
			\caption{Elastic ring model with geometric description}	
		\end{wrapfigure}	
		
		We consider a two-dimensional, thin, inextensible, uniform circular elastic ring in its equilibrium configuration. The ring deforms under the application of a uniform loading parameter $p$, which represents a constant inward pressure or compression. The curve $\Lambda$ describing the ring is parametrized by the arc length $s$, and its position vector is denoted as $r(s)=(x_1(s),x_2(s))$, where $x_1(s)$ and $x_2(s)$ are Cartesian coordinates in a Euclidean plane.
			
		The coordinates $x_1(s)$ and $x_2(s)$ can be expressed as integrals involving the angle $\vartheta(s)$, which represents the angle between the tangent to the curve and the x-axis:
		\begin{flalign*}
			x_1(s)&= \int \cos(\vartheta(s)) ds, \quad	x_2(s)= \int \sin(\vartheta(s)) ds,  \quad  0 \leq s \leq 2\pi.
		\end{flalign*}
		
		The angle $\vartheta(s)$ is given by:
		\begin{flalign*}
			\vartheta(s) = s + \int_0^s \nu(s) ds,
		\end{flalign*}
		where $\nu(s)$ represents the curvature difference between the deformed and undeformed ring.
		
		The Euler elastica model, derived using force and moment balance formulation, is given by:
		\begin{flalign}
			\nu^{\prime\prime}(s) + \mu \nu(s) - \beta + \frac{3}{2}\nu(s)^2 + \frac{1}{2}\nu(s)^3 = 0,
			\label{model}
		\end{flalign}
		where $\mu$ and $\beta$ are parameters related to the loading parameter $p$, with $p=\mu+\beta-1$.
		
		To ensure the closure of the ring, the following conditions must be satisfied:
		\begin{flalign}
			\nu'(0)=\nu'\left(\frac{\pi}{n}\right)=0, \quad \int_{0}^{\frac{\pi}{n}}\nu(s) ds=0,
			\label{condition1}
		\end{flalign}
		\begin{flalign}
			\nu(-s)=\nu(s), \quad \nu\left(s-\frac{2\pi}{n}\right) = \nu(s),
			\label{condition2}
		\end{flalign}
		where $n$ represents the number of-fold symmetry of the loading conditions.
		
		For more details on the formulation, refer to \cite{Tadjbakhs} and \cite{Flaherty}.
		\section{Method: Harmonic Balance}
		
		The harmonic balance method \cite{newh}, \cite{MI6C}, \cite{Nicolae} is employed to solve Eq. \eqref{model}. This method proposes a solution in the form of a Fourier series and determines the coefficients by balancing harmonics. The general expression for the Fourier series is given by:
		
		\begin{flalign*}
			\nu(s) = \sum_{k=0}^{\infty} (A_k \cos{(n  k  s)} + B_k \sin{(n  k s)}).
		\end{flalign*}
		
		To obtain the zeroth-order harmonic balance approximation for mod $n$, we start with an initial guess solution $\nu_0(s) = A \cos{(n s)}$, which satisfies the conditions \eqref{condition1} and \eqref{condition2}. Substituting this into Eq. \eqref{model}, we have:
		
		\begin{flalign}
			(A \cos{(n s)})^{\prime \prime} + \mu ( A \cos{(n s)}) - \beta + \frac{3}{2} (A \cos{(n s)})^2 + \frac{1}{2} (A \cos{(n s)})^3 = 0.
			\label{eq1}
		\end{flalign}
		
		By simplifying and balancing the harmonics in Eq. \eqref{eq1}, we solve for the required parameters, resulting in the zeroth-order approximation terms:
		
		\begin{flalign*}
			\nu_0(s) = A \cos{(n s)}, \quad \mu_0(A) = \frac{1}{8} \left(8 n^2-3 A^2\right), \quad \beta_0(A) = \frac{3}{4} A^2.
		\end{flalign*}
		
		For higher-order approximations, an iterative procedure is used with the following expression:
		
		\begin{flalign}
			\nu_k(s) = \nu_{k-1}(s) + \delta \nu_{k-1}(s), \quad \mu_k = \mu_{k-1} + \delta \mu_{k-1}, \quad \beta_k = \beta_{k-1} + \delta \beta_{k-1},
			\label{NL}
		\end{flalign}
		
		where $k$ represents the order of approximation, and $\delta \square$ denotes the incremental operator.
		
		The first approximation terms $v_1$, $\mu_1$, and $\beta_1$ are calculated using a similar procedure:
		
		\begin{flalign*}
			\nu_1 & = \frac{\left(3 A^3-6  A^2+8 A n^n-32 A  n^2\right)}{M} \cos (n  s)-\frac{6  A^2}{M} \cos (2  n  s), \\
			\mu_1 & = -\frac{9  A^4+24 A^2 \left(n^n-5  n^2-3\right)+48  A \left(n^n+n^2\right)-64  n^2 \left(n^n-4 n^2\right)}{8 \left(M\right)}, \\
			\beta_1 & = \frac{6  A^2 \left(n^n-4  n^2\right)}{M},
		\end{flalign*}
		
		where $M = 3  A^2-12  A+8 n^n-32  n^2$. A similar procedure is followed for higher-order approximations.
\section{Solution: Equilibrium Shapes}		
		Using the above formulation, we have computed the equilibrium shape of an elastic ring under loading conditions $n=2,3,4$ and $5$ as 	\cite{AMajid}. 
		\begin{figure}[H]
			\centering
			\begin{tabular}{cccc}
				\includegraphics[scale=0.275]{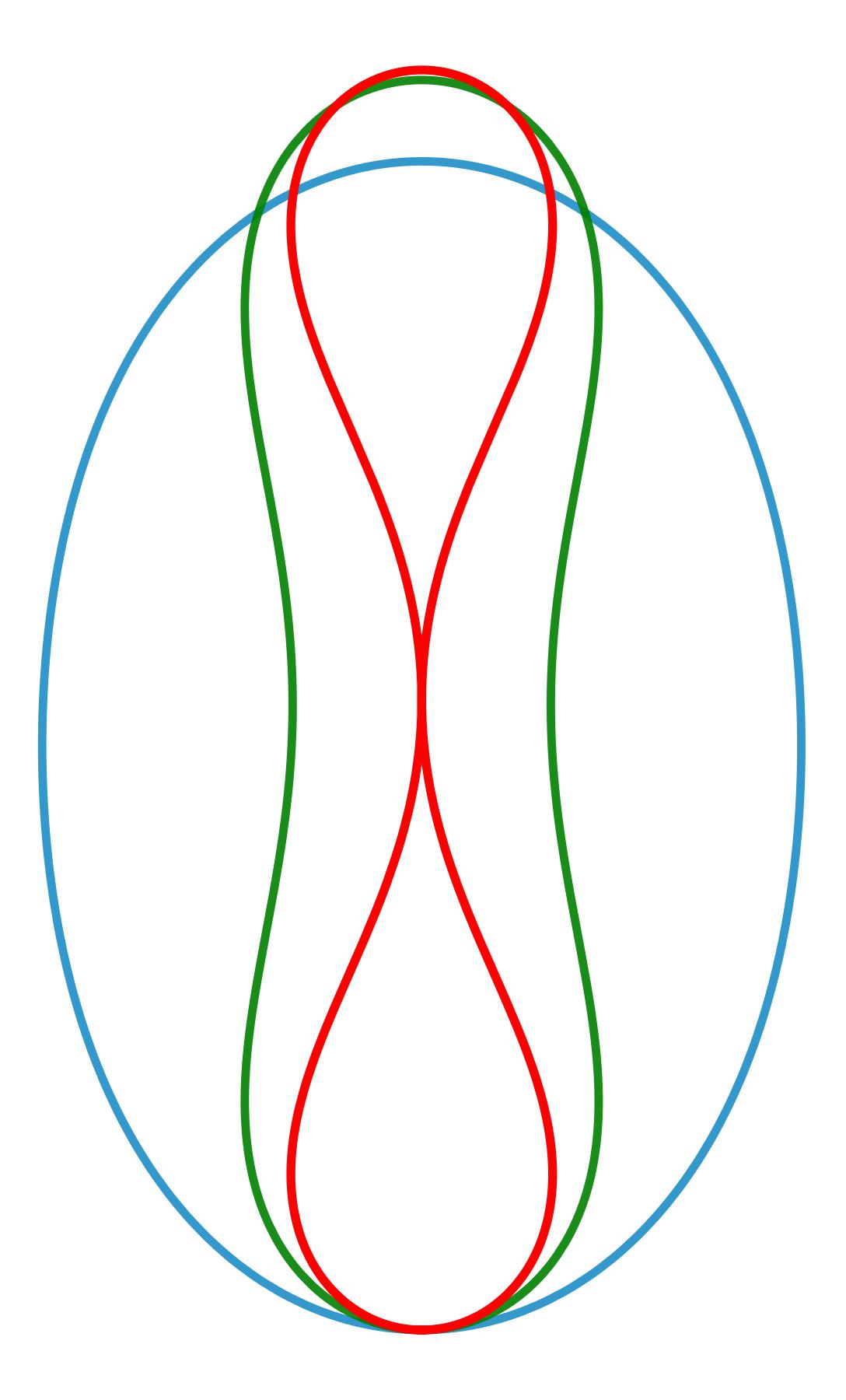} 
				\quad  
				& 
				\quad
				\includegraphics[scale=0.3]{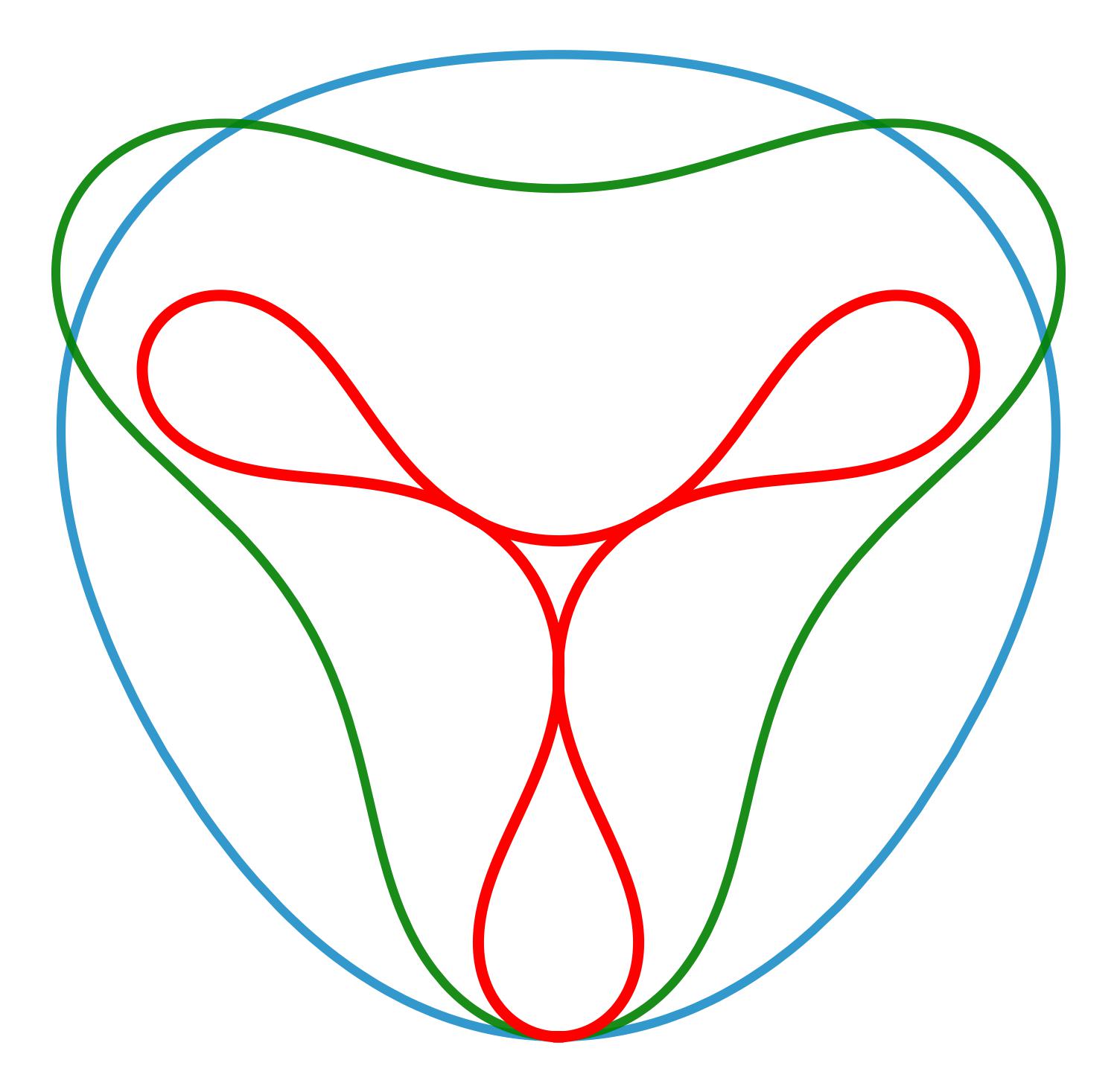}
				\quad
				&
				\quad
				\includegraphics[scale=0.3]{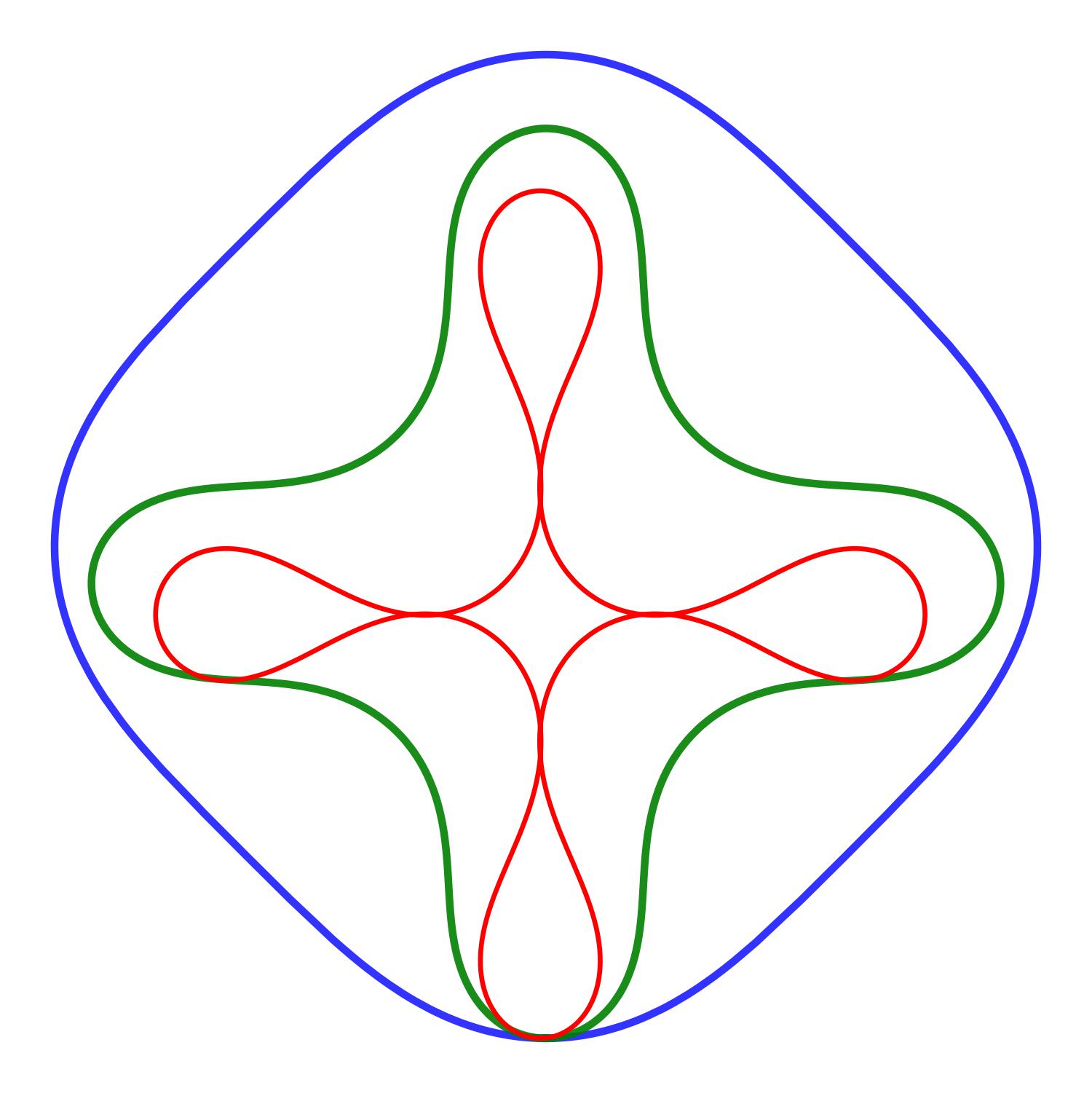}
				\quad
				&
				\quad
				\includegraphics[scale=0.3]{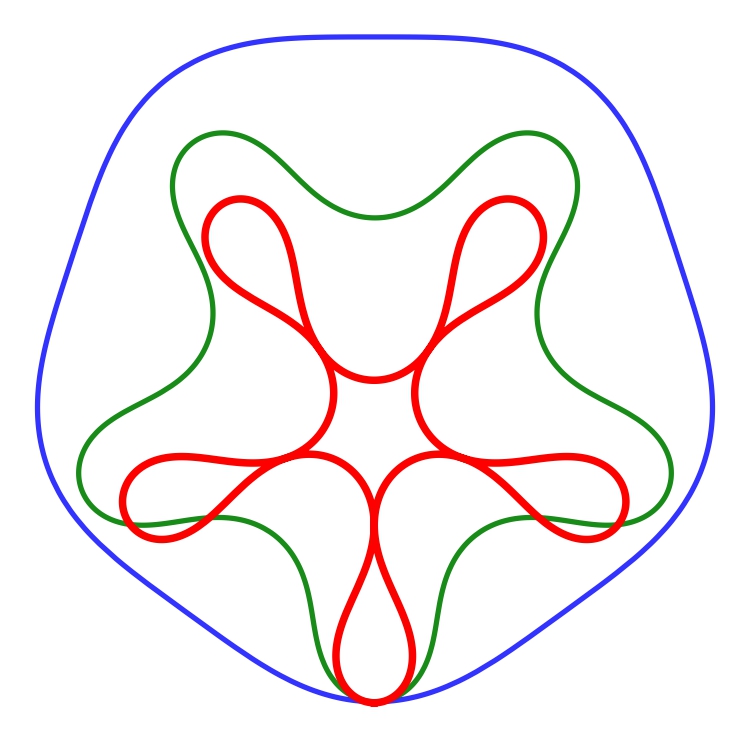}  
			\end{tabular}
			
			\label{figure}\caption{Equilibrium ring shapes  corresponding to (a) $p = 3, 4, 5.247$ (two fold contact shape); (b) $p =8, 15, 21.65$ (three fold contact shape); (c) $p = 15, 30, 51.844$ (four fold contact shape); (d) $p = 24, 77, 97.834$ (five fold contact shape)}
		\end{figure}
	As external loading move to the $p_{cp}<p_{si}$, so self-intersection start to occurring due to $2D$ modeling condition which induces the $x_2^- = x_1^+$ .Self intersection value and shape remained the point of interest for many researchers, because it is the point where physical system transform into different 
			
		\begin{figure}[H]
			\centering
			\begin{tabular}{cccc}
				\includegraphics[scale=0.08]{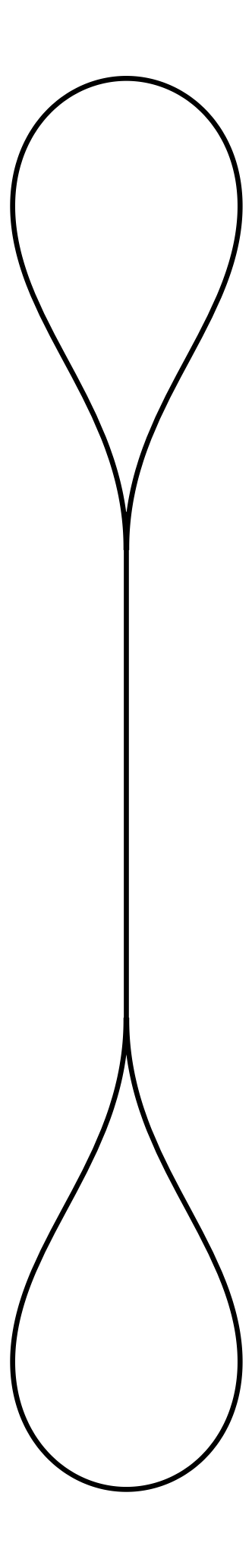}
				\qquad    
				& 
				\qquad
				\includegraphics[scale=0.12]{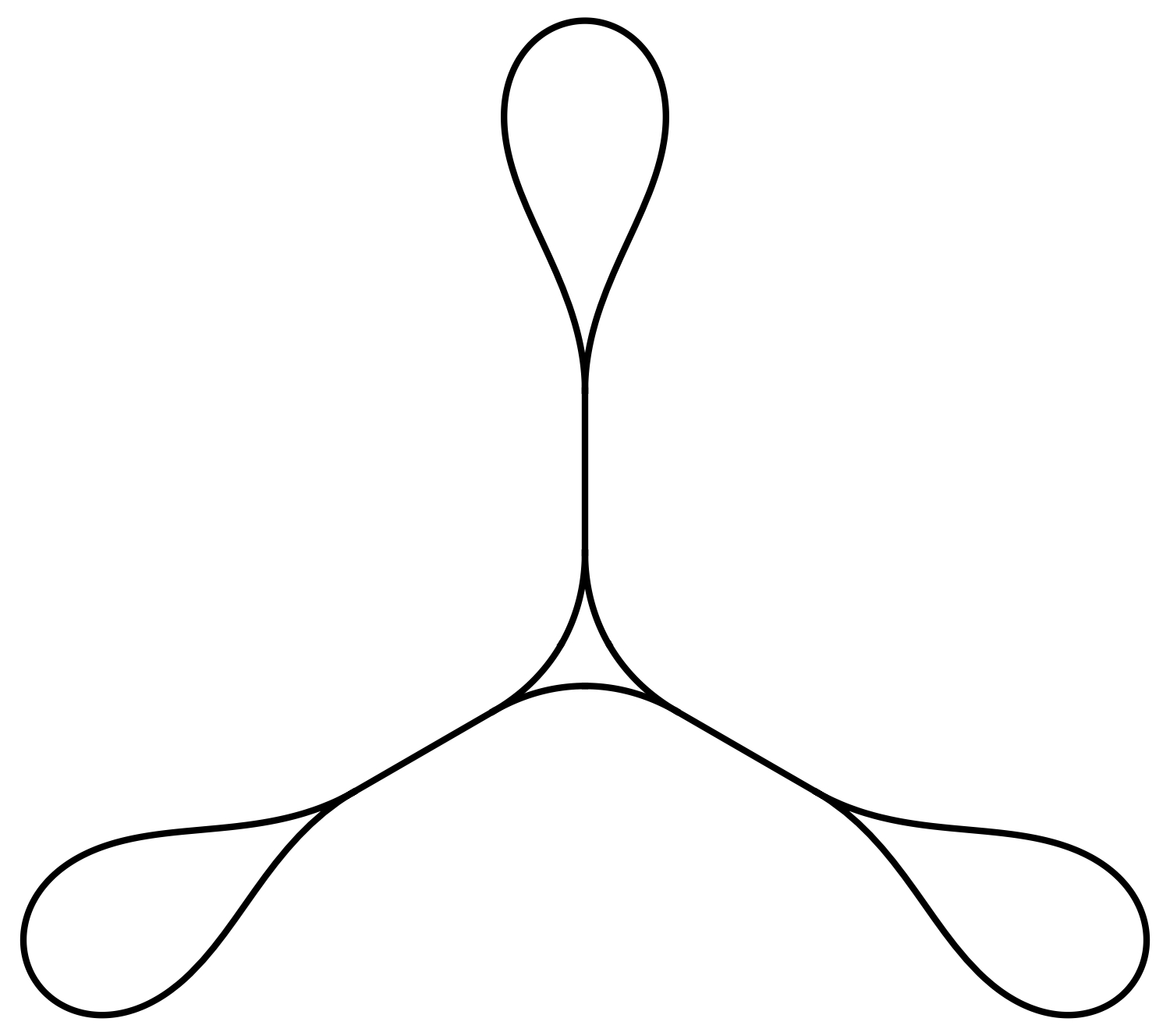}
				\qquad
				&
				\qquad
				\includegraphics[scale=0.13]{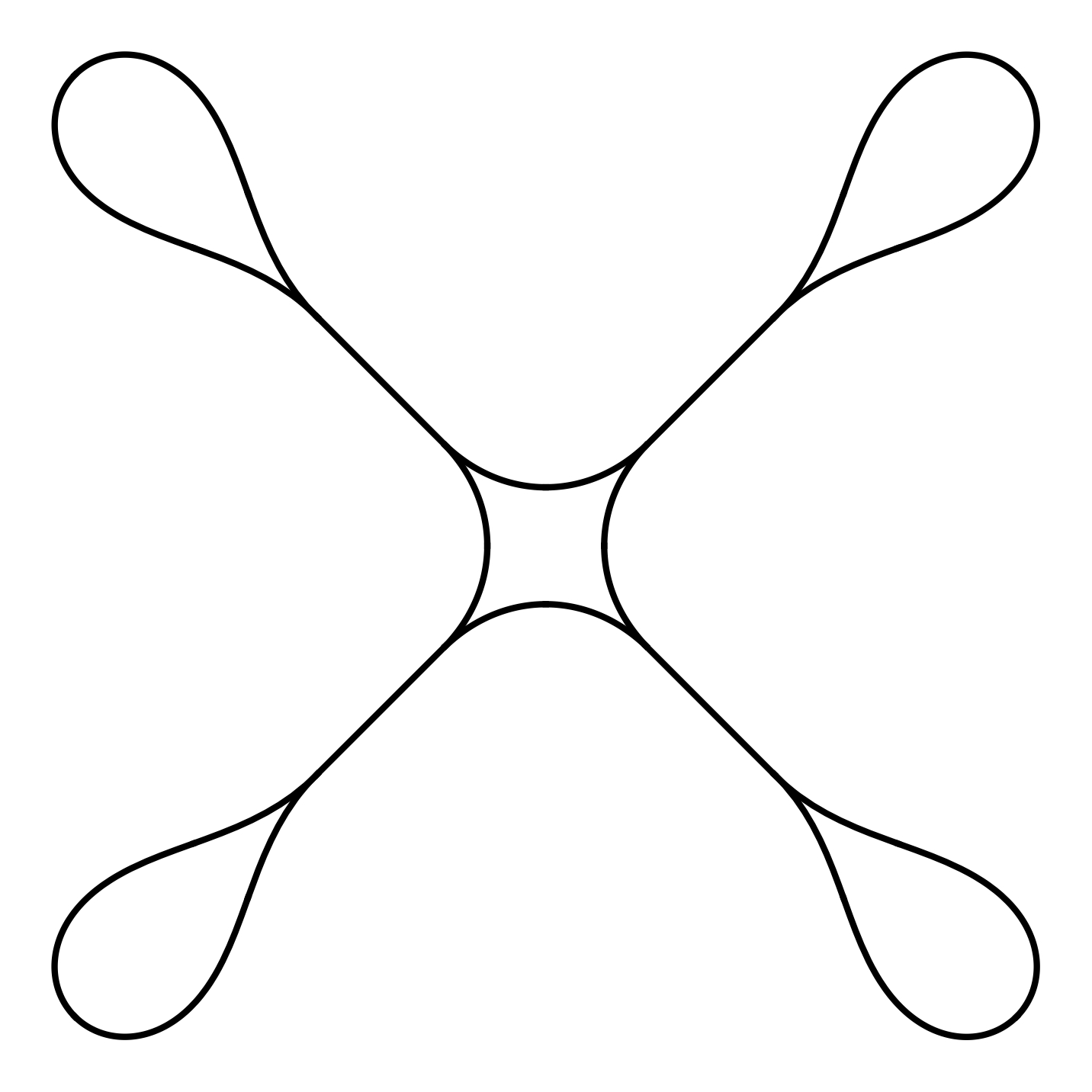}
					\qquad
				&
				\qquad
				\includegraphics[scale=0.13]{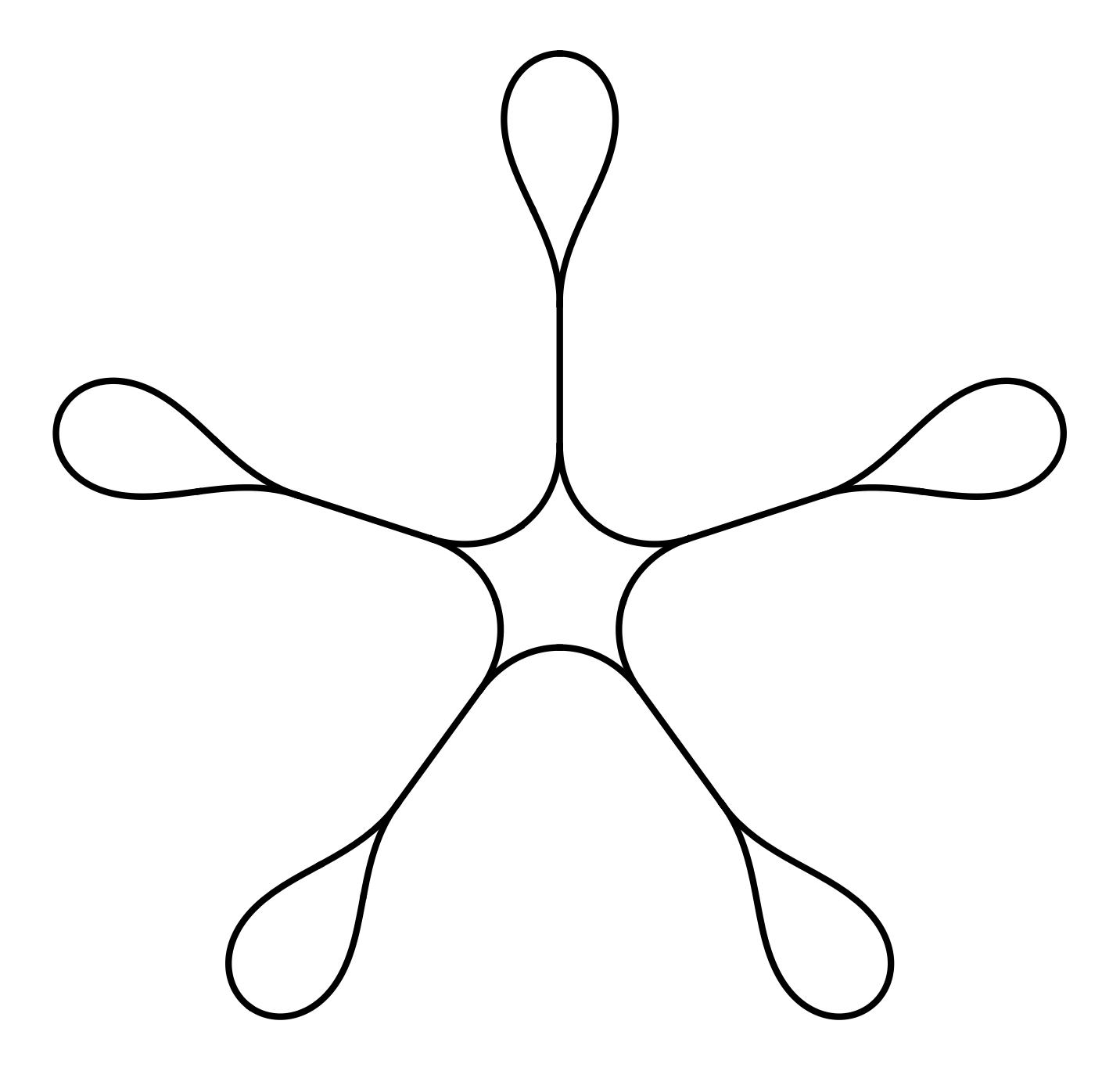}
			\end{tabular}
			
			\label{figure}\caption{Plot of $v^{2}(s)$ at $p=10.34$, $v^{3}(s)$ at $p=70.34$ and $v^{6}(s)$ at $p=600$}
		\end{figure}
		
		If we exceed the external loading, we would get intersecting shape because imposing the two-dimensional condition, and then this will reach to break-even point,where the stable shapes becomes unstable.

		\section{Stability Analysis}
		
		In our previous work \cite{AMajid2}, we used the third-order harmonic balance formulation to calculate the equilibrium shape of the Euler elastica. This formulation provided improved precision for all symmetric loading conditions. Therefore, we employ the same formulation to compute stability diagrams under the loading parameter $p$. The stability range varies with different loading conditions, starting from $n^2-1$ and bifurcating after a certain value of $p$.
		\begin{figure}[H]
			\centering
			\begin{subfigure}[b]{0.4\textwidth}
				\centering
				\includegraphics[width=\textwidth]{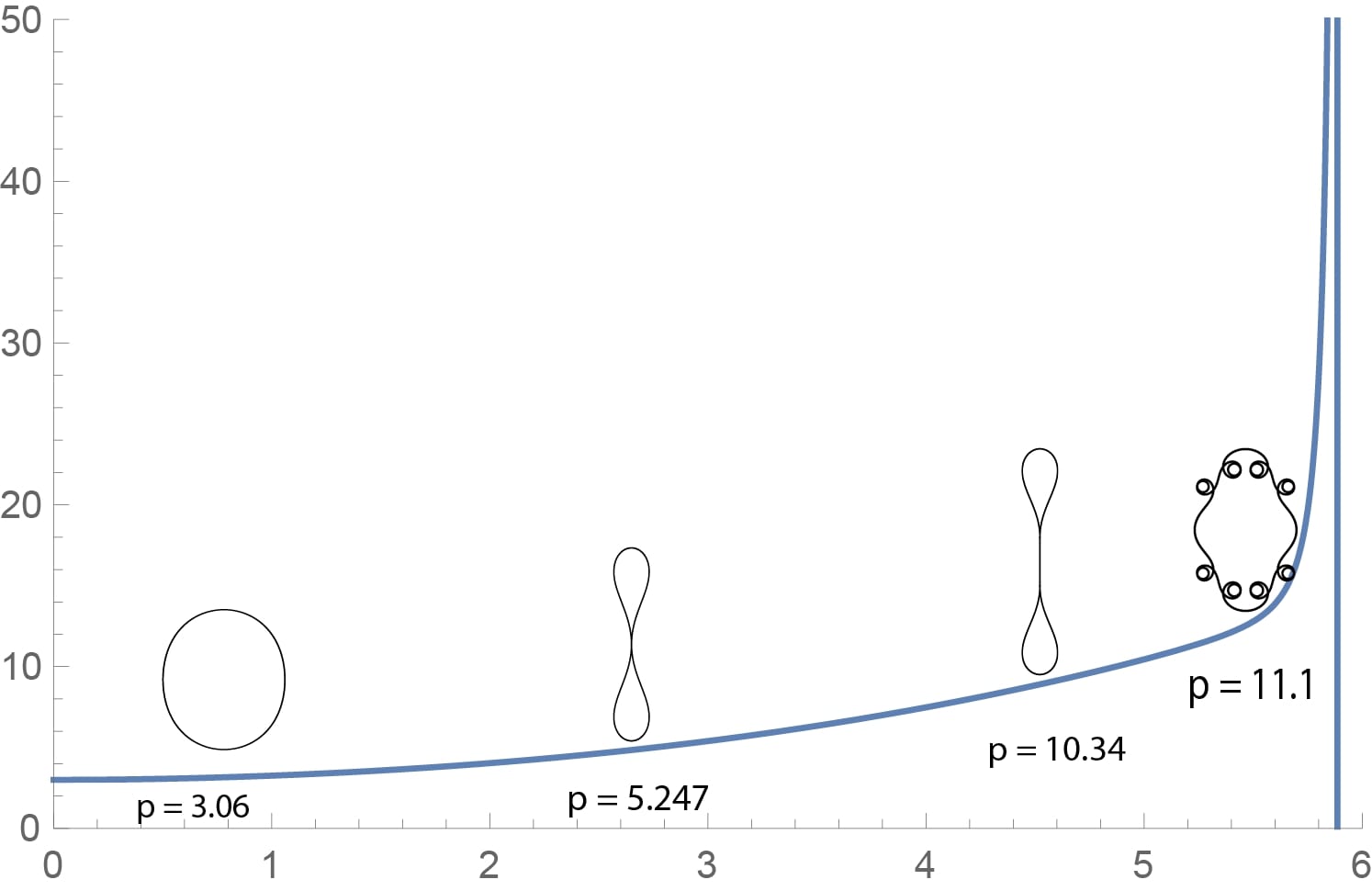}
				\caption{}
				\label{Pn2}
			\end{subfigure}
			\qquad
			\begin{subfigure}[b]{0.4\textwidth}
				\centering
				\includegraphics[width=\textwidth]{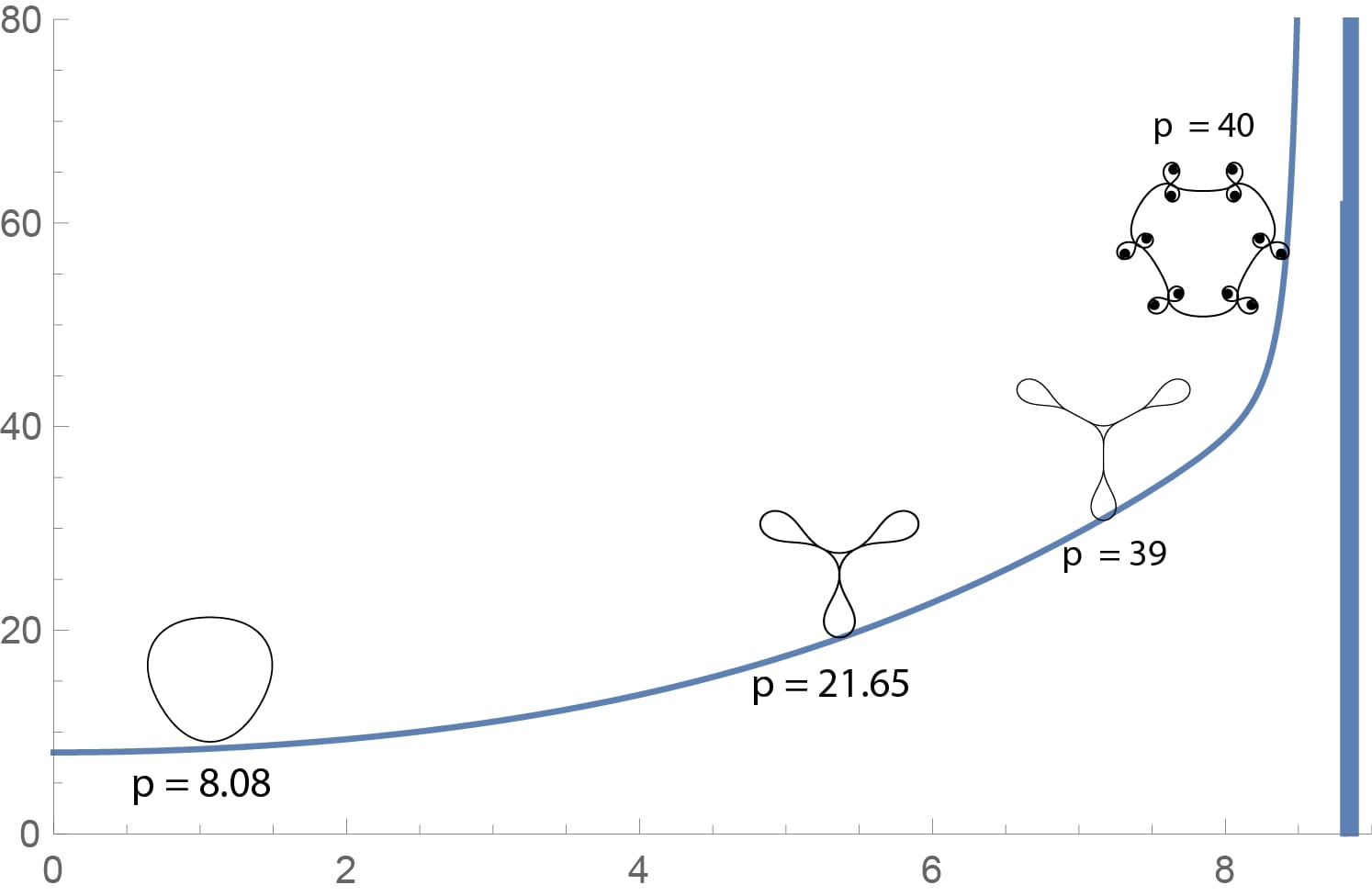}
				\caption{}
				\label{Pn3}
			\end{subfigure}
			\caption{Stability diagram between  $p$ vs. $\nu(0)$ at (a) $n=2$ and (b) $n=3$, with some stable and unstable elastic ring shapes at specific values of $p$}
			\label{Pplot1}
		\end{figure}

		\begin{figure}[H]
			\centering
			\begin{subfigure}[b]{0.4\textwidth}
				\centering
				\includegraphics[width=\textwidth]{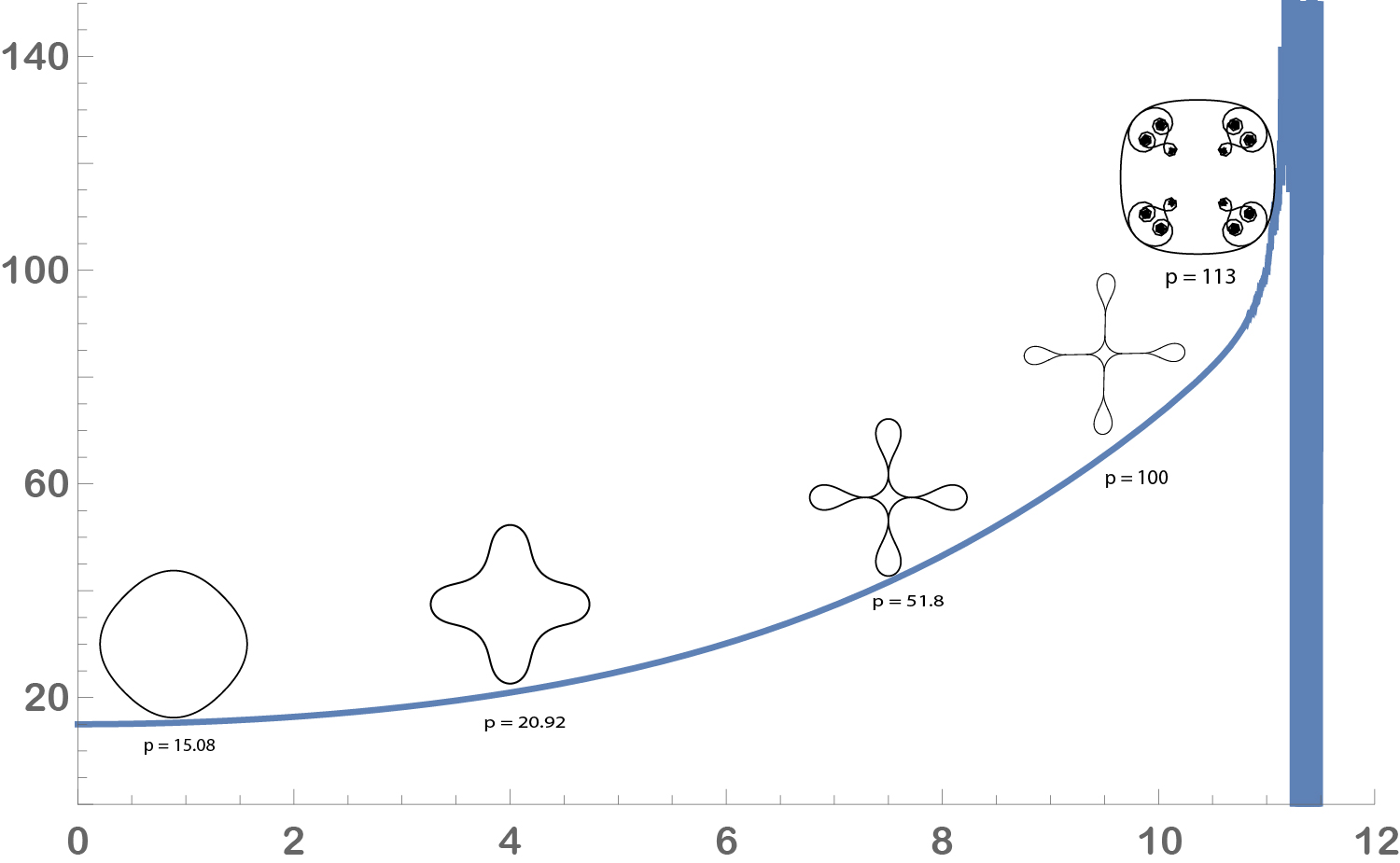}
				\caption{}
				\label{Pn4}
			\end{subfigure}
			\qquad
			\begin{subfigure}[b]{0.4\textwidth}
				\centering
				\includegraphics[width=\textwidth]{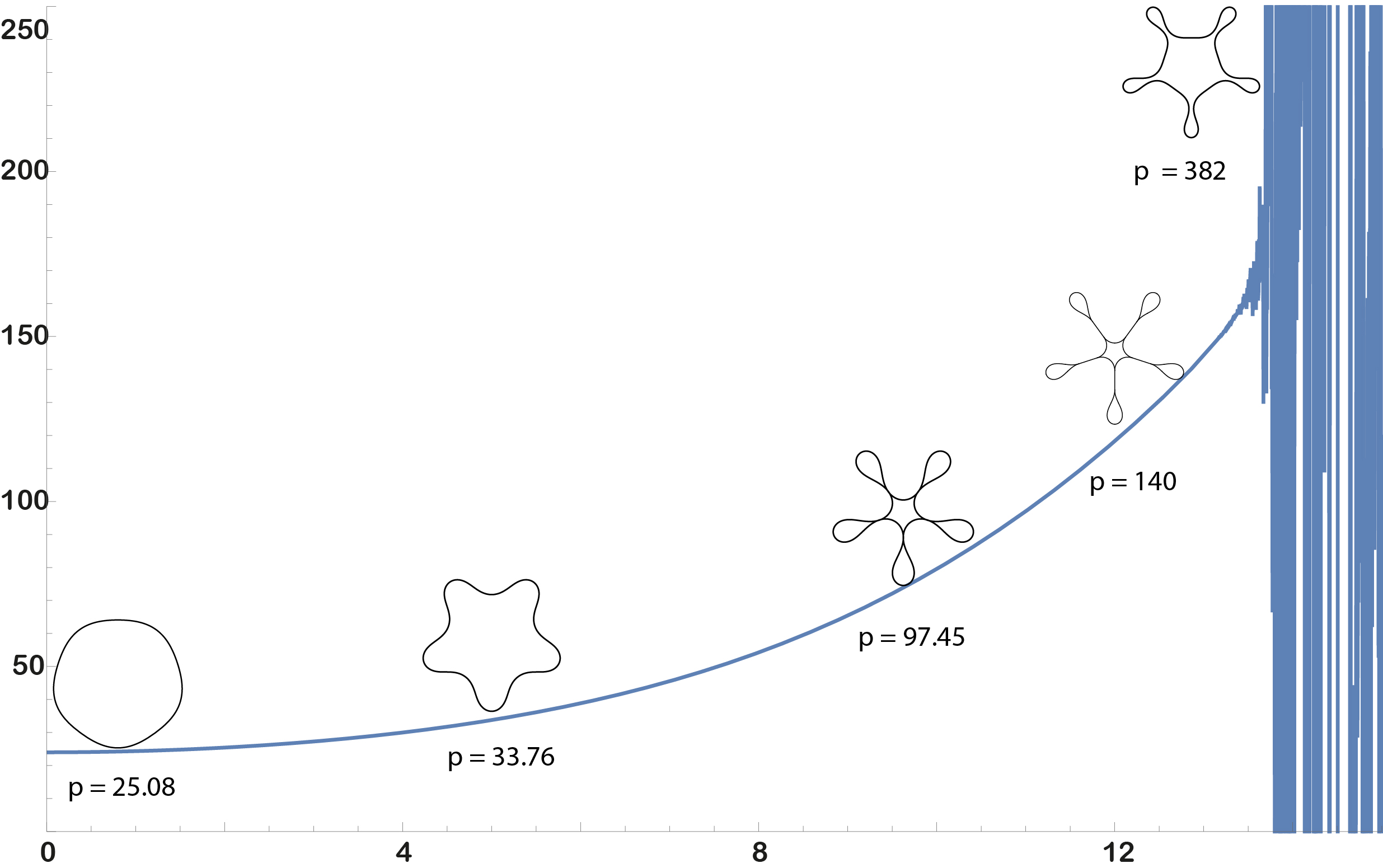}
				\caption{}
				\label{Pn5}
			\end{subfigure}
			\caption{Stability diagram between  $p$ vs. $\nu(0)$ at (a) $n=4$ and (b) $n=5$, with some stable and unstable elastic ring shapes at specific values of $p$}
			\label{Pplot2}
		\end{figure}
		Moreover, elastic ring shapes are computed over the stability diagram, which expresses the stable to unstable shape transition. The shapes in each of the graphs can be referred to (left to right) as closed, contact, over-contact, and irregular shape, as Fig. \ref{Pplot1}-\ref{Pplot2}.
		
		\begin{table}[htb]
			\centering
			\begin{tabular}{ |p{3cm}|p{3cm}||p{3cm}|p{3cm}|  }
				\hline
				\multicolumn{4}{|c|}{Stability Range by HBM} \\
				\hline
				Loading conditions $n$ & Stable Range $p$ &Loading conditions $n$ & Stable Range $p$ \\
				\hline
				2   & $(3,10.5)$    &7&   $(48,350)$ \\
				3 &    $(8,70)$ & 8   & $(63,500)$ \\
				4 & $(15,110)$ & 9 &  $(80,650)$ \\
				5   & $(24,170)$ & 10 &  $(99,799)$ \\
				6 &   $(35,250)$  & 12 & $(143,1191)$ \\
				\hline
			\end{tabular}
			\caption{ Stable range of $p$ for various loading conditions $n$ by using Newton--harmonic balance method.}
			\label{HBM_table}
		\end{table} 
		
		Figures \ref{Pplot1} and \ref{Pplot2} show the stability diagrams for different values of $n$, where $p$ is plotted against $\nu(0)$. The diagrams depict the transition from stable to unstable shapes, including closed, contact, over-contact, and irregular shapes at specific values of $p$.
		
		Table \ref{HBM_table} summarizes the stable ranges of $p$ for various loading conditions $n$ obtained using the harmonic balance method (HBM).
		
		While considering the numerical procedure(via Shooting technique) and computing through inbuilt numerical solving command $NSolve$ gives the same results. But while computing the equilibrium shape at certain value does not compute the shape, due to boundary conditions and equation compatibility. Here is the stability diagram at $n=5$.
			\begin{figure}[H]
			\centering
			{\includegraphics[width=0.45\textwidth]{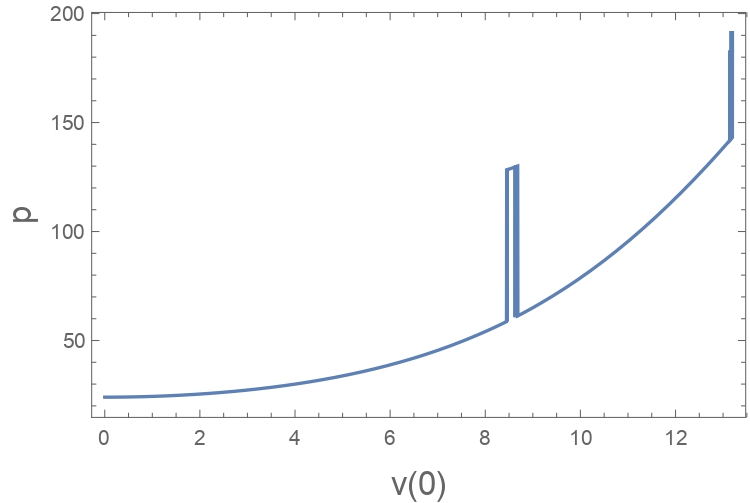}}
			\caption{Bifurcation diagram computed numerically at $n=5$.}
		\end{figure}
		
		Previous researcher have not computed the stability range and bifurcation diagrams of external loading/pressure $p$ using elliptic functions. Since elliptic functions are computed if the certain conditions computes at particular loading value $p*$,  and then find the equilibrium shape of ring as in ref{}.
		
		The stability ranges in each loading condition $n$ exhibit a unique pattern, starting from $n^2-1$. The computations were performed using Wolfram Mathematica \cite{Mathematica}, and the author can provide the Mathematica notebooks upon request.
		
		\section{Conclusion}
		
		In this paper, we have computed the stability ranges of the Euler elastica problem using the harmonic balance method. The obtained stability diagrams are smooth and extend to larger loading values compared to numerical solutions. The diagrams vary with the applied symmetric loading mod $n$, starting from a stable closed shape to a contact shape $p_c$, and then transitioning to a self-intersecting shape and an unstable region with irregular shapes.
		
		The harmonic balance method proves to be effective in computing stability diagrams for all loading conditions $n$. It demonstrates good agreement with the bifurcation theory of linearization, as the computed unstable ranges of the parameters $(\mu, \beta)$ align with the unstable regions in the harmonic balance results.
		
		In conclusion, the harmonic balance method provides accurate stability calculations for the Euler elastica, yielding stable equilibrium shapes over an extended range of the loading parameter $p$. It also aligns well with the bifurcation theory of linearization, indicating the unstable regions in the parameter space
		
		\section*{Declaration of competing interest}

		The authors disclose that they have no financial or non-financial interests directly or indirectly related to the work submitted for publication.

	\end{document}